\documentclass[final,5p,times,twocolumn]{elsarticle}
\usepackage{amssymb}
\usepackage{amsmath}

\begin{document}

\begin{frontmatter}

\title{One- and two-channel
Kondo model with logarithmic Van Hove singularity: a numerical
renormalization group solution
}

\ead{Zhuravlev@imp.uran.ru, Valentin.Irkhin@imp.uran.ru}

\author{A. K. Zhuravlev,
A. O. Anokhin and
V. Yu. Irkhin}


\address{M.N. Mikheev Institute of Metal Physics, Russian Academy of Sciences, 620108 Ekaterinburg, Russia}

\begin{abstract}
Simple scaling consideration and NRG solution of the one- and two-channel Kondo model in the presence of a
logarithmic Van Hove singularity at the Fermi level is given. The temperature dependences of local and impurity
magnetic susceptibility and impurity entropy are calculated. The low-temperature behavior of the impurity
susceptibility and impurity entropy turns out to be non-universal in the Kondo sense and independent of the $s-d$ coupling~$J$.
The resonant level model solution in the strong coupling regime  confirms the NRG results.
In the two-channel case the local  susceptibility demonstrates a non-Fermi-liquid power-law behavior.

\end{abstract}





\end{frontmatter}

Anomalous $f$- and $d$-systems possess highly unusual electronic properties and magnetism. Besides the heavy-fermion
behavior, they demonstrate the non-Fermi-liquid (NFL) behavior: logarithmic or anomalous power-law temperature dependences
of magnetic susceptibility and electronic specific heat~\cite{Stewart1l}. Their magnetism has both localized and itinerant
features, being determined by both Kondo effect and density of states (DOS) singularities which are especially important
for magnetic ordering.

The NFL behavior is related to peculiar features of electron and spin fluctuation spectra.
 In particular, the multichannel Kondo model is often used which assumes existence of degenerate electron
bands. This model explains power-law or logarithmic behavior of electronic specific heat and magnetic susceptibility~\cite{Wiegmann,Tsv,Cox}.
Recently, the one- and two-channel charge Kondo effect was extensively discussed for
nanostructures, layer systems and quantum dots~\cite{Nature,Law1,Khar,Pot1}.

In the present Letter we treat the Kondo model with the electron spectrum containing a logarithmic DOS singularity. This
van Hove singularity is typical, in particular, for the 2D case.
This is present, e.g., in the electron spectrum of graphene where
the Kondo effect is tunable with carrier density~\cite{Chen}. Recently, a possibility of graphene doping
for the exploration of Van Hove physics was proposed~\cite{Takeshi}.

Whereas the flat-band Kondo model permits exact Bethe ansatz solution, the model with singular DOS is a challenge for
analytical field-theoretical consideration. At present, the case of empty conduction band was investigated by numerical
renormalization group (NRG)  method~\cite{Zitko}.
A NRG and  resonant level model treatment of the power-law and $1/|E|\ln^2|E|$
divergent bare DOS was performed in Refs.\cite{Mitchell}.
To investigate the case of singular DOS for the Kondo metal we use a
simple scaling consideration and compare the results with more advanced NRG calculations.

We start from the Hamiltonian of the one-center $s-d(f)$ exchange (Kondo) model
\begin{equation}
H_{sd}=\sum_{\mathbf{k} m \alpha }\varepsilon _{\mathbf{k}}c_{\mathbf{k} m \alpha
}^{\dagger }c_{\mathbf{k} m \alpha }^{{}}-\sum_{\mathbf{k}\mathbf{k^{\prime }}%
m\alpha \beta }J_{\mathbf{k}\mathbf{k^{\prime }}}\mathbf{S}%
\mbox {\boldmath
$\sigma $}_{\alpha \beta }c_{\mathbf{k}m\alpha }^{\dagger }c_{\mathbf{%
k^{\prime }}m\beta }^{{}}\ .  \label{Ham_Kondo}
\end{equation}%
Here $\varepsilon _{\mathbf{k}}$ is the band energy, $\mathbf{S}$ are spin operators with spin value being $S$,
$\mathbf{\sigma}$ are the Pauli matrices, in the case of contact coupling $J_{\mathbf{k}\mathbf{k^{\prime }}}=J/N_{s}$
where $J$ is the $s-d(f)$ exchange parameter, $N_{s}$ is the number of lattice sites, $m=1...M$ is the orbital degeneracy
index, $\alpha, \beta$ are spin indices.

The density of states corresponding to the spectrum $\varepsilon_{\mathbf{k}}$ is supposed to contain a Van Hove
singularity near the Fermi level. In particular, for the square lattice with next- and next-to-nearest neighbour transfer
the spectrum reads
\begin{equation}
\varepsilon_{\mathbf{k}}=2t(\cos k_{x}+\cos k_{y})+4t^{\prime }(\cos k_{x}\cos k_{y}+1)
\end{equation}
and
we have the density of states%
\begin{equation}
\rho (E) \simeq \frac{1}{2\pi ^{2}\sqrt{t^{2}-4t^{\prime 2}}}\ln \frac{16%
\sqrt{t^{2}-4t^{\prime 2}}}{|E|}
\label{Eq:rho0}
\end{equation}
where
the bandwidth is determined by $|E-8t^{\prime }|<4|t|$.
Below we use the approximate density of states for  $t^{\prime }=0$
\begin{equation}
 \rho (E)=\varrho F(E), \ F(E)=\ln \frac{4D}{|E|} , \ \varrho=\frac{2}{\pi^2 D},\
 D= 4t \ .
 \label{Eq:rho}
\end{equation}

We apply the \textquotedblleft poor man scaling\textquotedblright\ approach~\cite{And}.
This considers the dependence of
effective (renormalized) coupling  $J_{ef}(C)$ on the flow cutoff parameter $C\rightarrow -0$ which occurs at picking out
the singular Kondo contributions.

To find the scaling equation we pick out in the sums for the Kondo terms the contribution of intermediate electron states
near the Fermi level with $C<\varepsilon_{\mathbf{k}}<C+\delta C$ to obtain to next-leading order in $J$ (see details in~\cite{I11,I16})
\begin{equation}
\delta J_{ef}(C)=2\varrho J^{2}[F(C) + J\varrho MF(C/2)F(-C/2)]\delta C/C
\label{ief}
\end{equation}%
where the factor of $F$ comes from the singularity of DOS.
For the correction to the localized magnetic moment $S_{ef}$ determining the Curie constant,
as determined from the Kondo contribution to local magnetic susceptibility~\cite{Kondo,IK89},
we have (cf.~\cite{I11})%
\begin{equation}
\delta S_{ef}(C)/S=2\varrho ^{2}J^{2}MF(C/2)F(-C/2)\delta C/C \ .
\end{equation}

The lowest-order scaling calculation according to~(\ref{ief})  (see also Refs.~\cite{Zhur2009,I11}) yields for the boundary
of the strong-coupling region
\begin{equation}
T_{K}\varpropto D\exp \left[ -\left( \frac{\pi ^{2}D}{2|J|}\right) ^{1/2} \right] .  \label{perturb}
\end{equation}

However to describe a possible NFL-type behavior (intermediate-coupling fixed point), we have to use the next-order scaling equations
\begin{equation}
\frac{\partial g_{ef}(\xi )}{\partial \xi} =[\xi-\frac{M}{2}(\xi +\ln 2)^2 g_{ef}(\xi )]g_{ef}^{2}(\xi ) \label{sc2}
\end{equation}%
where $\xi =\ln |4D/C|$ and we have  introduced the dimensionless
effective  $s-d$ coupling constant%
\begin{equation}
g_{ef}(C) = -2\varrho J_{ef}(C) \ . 
\end{equation}%
In the flat-band case (where the singular factors $\xi +\ln 2$ are replaced by unity) such equations give a finite fixed point  $g_{ef}(\xi\rightarrow \infty)=2/M$.
It is known that this point is unphysical (unreachable) for $M=1$, but for $M>2$ the scaling
consideration gives a qualitatively correct description, see review paper~\cite{Cox}. The case $M=2$ is marginal, so that
logarithmic factors occur which are missed by simple approaches.

Unlike the flat-band case, the equation~(\ref{sc2}) cannot be solved analytically, but only asymptotic solution
for $G_{ef}(\xi )\equiv g_{ef}(\xi )\xi$  at large
$\xi $ can be obtained,
which has the form%
\begin{equation}
G_{ef}(\xi )
=\frac{2}{M}+\left(1-\frac{4\ln 2}{M}\right)\frac{1}{\xi} \ .
\label{sc3}
\end{equation}
The second term can change sign and is positive for large $M$, so that the derivative of $G_{ef}(\xi)$ changes its sign.
Thus the details of scaling behavior are rather sensitive to parameters. Besides that,
the factors in~(\ref{sc2}) is well
determined only within the $1/M$-expansion. Nevertheless, these results demonstrate existence of the
``fixed point'' $ G_{ef}(\xi )\rightarrow 2/M$

By analogy with Refs.~\cite{I11,I16} we can write down the scaling equation
for the effective localized magnetic moment%
\begin{equation}
\frac{\partial \ln S_{ef}(\xi )}{\partial \xi} =-\frac{M}{2}G_{ef}^{2}(\xi )
\end{equation}
so that to leading approximation%
\begin{equation}
S_{ef}(C)\simeq (|C|/T_{K})^{\Delta },\Delta =2/M
\end{equation}
and for the local magnetic susceptibility,
\begin{equation}
\chi _{\mathrm{loc}}(T)=\int\limits_{0}^{1/T}\langle S_{z}(\tau
)S_{z}\rangle d\tau \ ,  \label{ChiLocal}
\end{equation}%
we obtain  the power-law dependence
\begin{equation}
\chi_{\mathrm{loc}} (T)\varpropto S_{ef}^{2}(T)/T\varpropto (T/T_{K})^{2\Delta -1}
\label{susc}
\end{equation}
where we have taken into account that $G_{ef}(C)$ reaches the value about $ 2/M$ at $|C| \sim T_K$. The dependence
$\chi_{\mathrm{loc}}(T)$ follows at high temperatures the Curie--Weiss law, has a maximum at $T \sim T_K$ and decreases
with further increasing $T$ for small $M$ (however, in this case the scaling results for $T <T_K$ are not reliable). For
large $M$, $\chi_{\mathrm{loc}}(T)$ is divergent at low temperatures.

 Note that the exponent in~(\ref{susc}) becomes modified in higher orders in $1/M$, and in the flat-band case one has
$\Delta =2/(M+2)$ according to the Bethe ansatz solution, see Refs.~\cite{Gan,Cox}. Below we take into account this
replacement at comparison of analytical and NRG results.

We see that simple analytical methods do not provide definite results, which is connected with insufficient information
provided by $1/M$ expansion. In particular the result for susceptibility~(\ref{susc})
does not differ from the
corresponding flat-band result and does not take into account the logarithmic factor at $M=2$~\cite{Wiegmann,Cox}.

At the same time a simple ``poor man'' renormalization group treatment captures the differences in the perturbation
expansion for the singular DOS and flat-band cases. Leaving the algebraical structure of the perturbation series the same,
it leads to the expansion in terms of $G_{ef}(\xi)$ for the singular DOS rather than $g_{ef}(\xi)$ for the flat-band case.
Moreover, it gives the inverse logarithmic contributions to the impurity entropy and the specific heat (see below).

Therefore, we calculate impurity magnetic susceptibility, entropy and specific heat by using numerical renormalization
group (NRG) approach~\cite{Wilson} in the one- and two-channel cases.

\begin{figure}[htbp]
\includegraphics[width=0.8\columnwidth, angle=0]{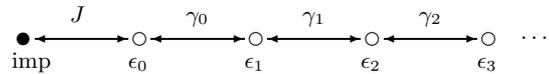}
\caption{Representation of the Kondo model in the form of a semiinfinite Wilson chain} \label{Fig_chain}
\end{figure}

\begin{figure}[htbp]
\includegraphics[width=0.8\columnwidth, angle=0]{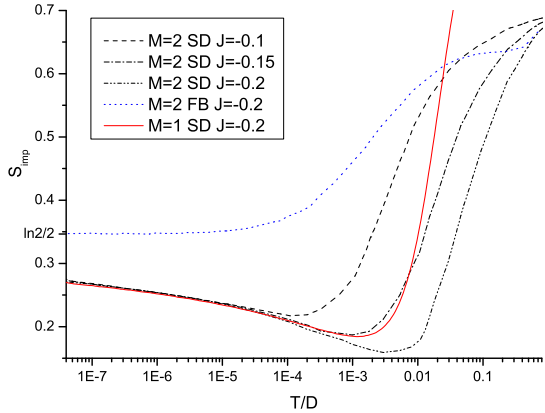}
\caption{NRG results for the impurity entropy $S_{\mathrm{imp}}$ in the two-channel case for singular DOS (SD) and non-singular DOS (FB, flat band) logarithmic
singularity,  and for the one-channel case with singular DOS. The latter curve is shown with the shift by $\ln 2 /2$.} \label{Fig:entropy_combine}
\end{figure}

The NRG procedure starts from the solution of the isolated-impurity problem
(sites ``imp''
and $\epsilon _{0}$ in Fig.~\ref{Fig_chain}).
At the initial step, we add a first conducting electronic site $\epsilon _{1}$, and construct and diagonalize a Hamiltonian
matrix on this Hilbert space (with a $4^M$--fold higher dimensionality). This procedure is multiply
repeated. However, since the dimensionality of the Hilbert space grows as
$4^{MN}$ ($N$ is the number of an iteration), it is impossible to store all the eigenstates during the calculation.
Therefore, it is necessary to retain after each iteration only the states with the lowest energies. If we restrict
ourselves to a certain maximum number of stored states (determined by the computational possibilities), it is necessary,
starting from a certain iteration, to retain of the order of $1/4^M$ of states at each step. Practically, the number of the
states is reasonable in the one- and two-channel cases. To take into account the disturbance introduced by the elimination
of the high-lying states we use Wilson's logarithmic discretization of the conduction band~\cite{Wilson} (see also Ref.~\cite{Zhur2009}).
In real calculations for $M=2$ we took into account of the order of $10^4$ states at each NRG
iteration with Wilson's logarithmic discretization factor being $\Lambda=3$. The agreement in the entropy value $S(T=0)$
with the Bethe ansatz results for the flat-band DOS, which is $\ln 2/ 2$,~\cite{Cox} (see below
Fig.~\ref{Fig:entropy_combine}) demonstrates sufficient of NRG calculations. For the case $M=1$ we performed NRG
calculations using $\Lambda=2$ and $\Lambda=1.5$ with subsequent extrapolation to $\Lambda=1$ which corresponds to non-discretized zone.

When performing NRG calculations in the singular DOS case, we chose $t=1/4, t^{\prime}=0 $
in approximation~(\ref{Eq:rho0})
and calculated   the half-width of the DOS support $D^{\prime}$ from the normalization condition, which gave $D^{\prime}=1.05977 D$.



\begin{figure}[htbp]
\begin{tabular}{c}
\includegraphics[width=0.8\columnwidth, angle=0]{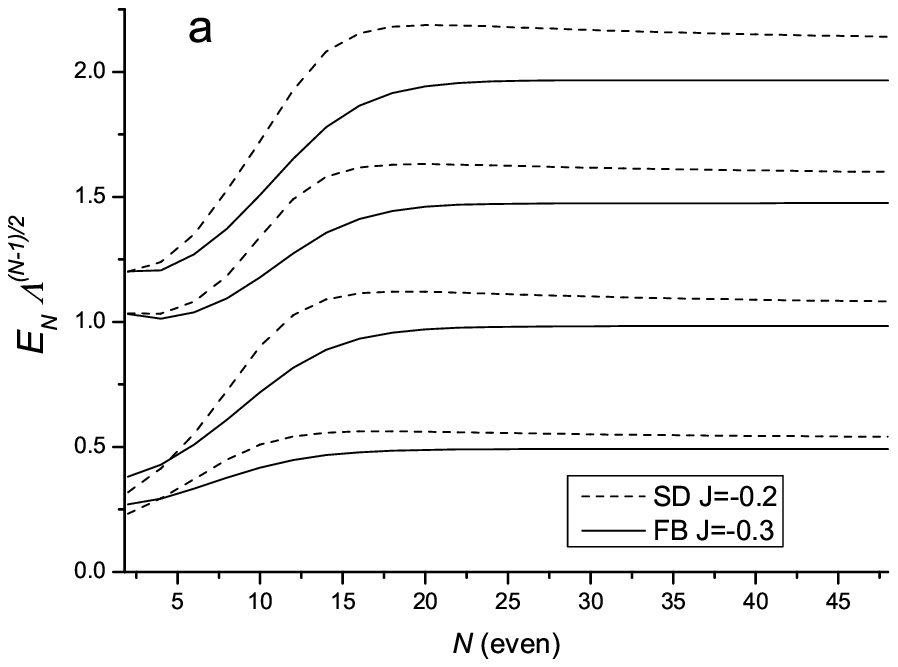} \\
\includegraphics[width=0.8\columnwidth, angle=0]{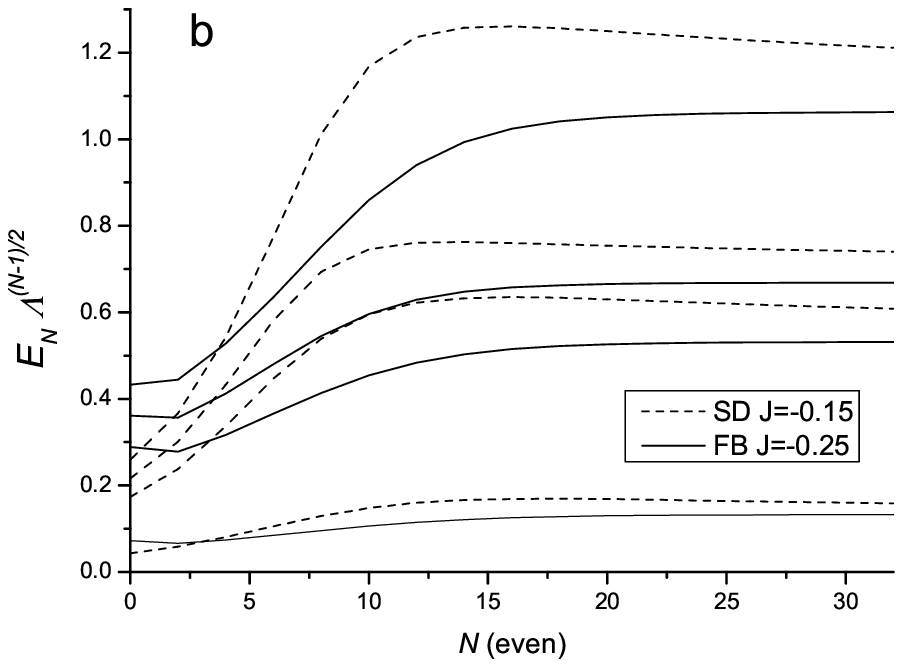} \\
\end{tabular}
\caption{The picture of lowest energy levels for flat-band~(FB) DOS and for DOS with Van Hove singularity~(SD): (a) the
one-channel model, $\Lambda=2$, (b) the two-channel model, $\Lambda=3$} \label{Fig:levels}
\end{figure}


The picture of energy levels depending on the NRG step $N$ (multiplied by $\Lambda^{(N-1)/2}$)
is presented in
Figs.~\ref{Fig:levels}. The energies are referred to the energy of the ground state. For flat-band and singular DOS cases
the values of the corresponding parameters $J$ were chosen from the approximate equality of Kondo temperatures $T_K$ for
the cases addressed. An important difference between the flat-band and singular DOS situations is considerably slower
tending of the curves $E(N)$ to the asymptotic values in the latter case (cf. Ref.~\cite{NRG}).
The slow fall off in the Van Hove case is somewhat
similar to the situation for an underscreened Kondo model \cite{Mehta}.

Because of  retaining only part of the energy spectrum at the $N$-th step of the NRG procedure,
thermodynamic averages should be calculated at a temperature that depends on
$\Lambda$, $T_N = \Lambda^{-N/2}T_0$ \cite{Wilson}.
Here the starting temperature $T_0$ should be  not too small to avoid the problem of discreteness
of the energy spectrum. The total entropy and  specific heat read \cite{Pruschke}
\begin{equation}
\mathcal{S}_{\mathrm {tot}} = \langle H\rangle_{\rm tot}/T + \ln Z_{\rm tot} , \
C_{\rm tot} = \left[\langle H^2\rangle_{\rm tot}-\langle H\rangle_{\rm tot}^2\right]/T^2 , \label{SCtot}
\end{equation}
where $Z$ is partition function. On differentiating $\langle S_z\rangle_{\rm tot}$ with respect to magnetic field one obtains \cite{Wilson}
\begin{equation}
T\chi_{\rm tot}(T) = \langle S_z^2\rangle_{\rm tot} - \langle S_z\rangle^2_{\rm tot} \ , \label{Tchitot}
\end{equation}
The quantities $T\chi_{\rm band}(T)$, $\mathcal{S}_{\rm band}$ and $C_{\rm band}$  are calculated in a similar way, and the
corresponding impurity contributions are obtained by subtracting them from (\ref{SCtot})-(\ref{Tchitot}).

First we consider the results for $M=1$ and discuss the local magnetic
susceptibility $\chi_{\mathrm{loc}}$ (\ref{ChiLocal}).
This is the susceptibility of a single impurity in a magnetic field that acts locally only on this impurity; this can be
measured experimentally from the impurity spin correlation function and is obtained in simple perturbation calculations~\cite{Kondo}.

Instead of calculating~(\ref{ChiLocal}) directly we use the following procedure. One can apply small magnetic field $h$ to
impurity spin only and then calculate numerically derivative
of the local magnetization induced, $d\langle S_z\rangle/dh$,
in the limit $h\to 0$. To be sure that this limit with a linear dependence $\langle S_z\rangle\propto h$ has been reached
we performed calculations for a series of $h$ values.

\begin{figure}[htbp]
\includegraphics[width=0.8\columnwidth, angle=0]{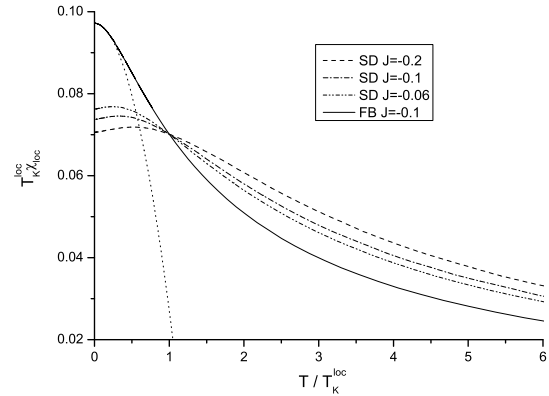}
\caption{NRG results for $T_K^{\mathrm{loc}}\chi _{\mathrm{loc}}$ extrapolated for $\Lambda=1$ for singular DOS~(SD) and
for flat-band case~(FB) with the corresponding parabolic fitting (dotted line) for one-channel case.}
\label{Fig_chi_loc_1chan_parabola}
\end{figure}

\begin{figure}[htbp]
\includegraphics[width=0.8\columnwidth, angle=0]{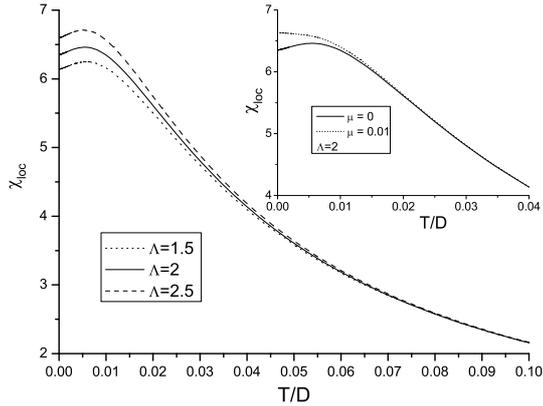}
\caption{NRG results for $\chi _{\mathrm{loc}}$. One-channel case with singular DOS, $J=-0.2$, the value of $\Lambda$ varies. $\mu$ is the distance between singularity and the Fermi level. }
\label{Fig_chi_loc_1chan}
\end{figure}

\begin{figure}[htbp]
\includegraphics[width=0.8\columnwidth, angle=0]{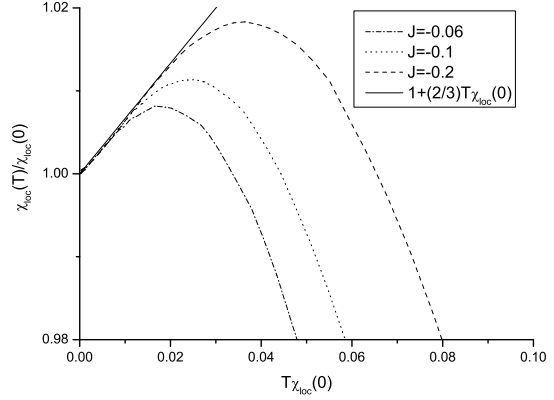}
\caption{NRG results for $\chi _{\mathrm{loc}}$ extrapolated to $\Lambda=1$ and the results of linear fitting. One-channel
case with singular DOS.} \label{Fig_chi_loc_TL}
\end{figure}

Numerical results for $\chi _{\mathrm{loc}}$ are shown in
Figs.~\ref{Fig_chi_loc_1chan_parabola}--\ref{Fig_chi_loc_TL}.
They clearly demonstrate that there exists characteristic crossover temperature $T_{K}^{\mathrm{loc}}$ for $\chi_{\mathrm{loc}}$.
Similar to Wilson~\cite{Wilson} we use (somewhat ambiguously) the
definition $T_{K}^{\mathrm{loc}}\chi_{\mathrm{loc}}(T_{K}^{\mathrm{loc}})=0.0701$
(see also Ref.~\cite{ZhurKett} and
discussion therein). The values of $T_{K}^{\mathrm{loc}}$ are presented in Table~\ref{Tabl:TK}.

However, unlike flat-band case, we did not observe exact universal behavior of $T_{K}^{\mathrm{loc}}\chi_{\mathrm{loc}}
(T)$ as a function of $T/T_{K}^{\mathrm{loc}}$ (see Fig.~\ref{Fig_chi_loc_1chan_parabola}).

\begin{table}[htb]
\caption{NRG calculations for $T_{K}^{\mathrm{loc}}$ with $J$ varying }
\begin{center}
\begin{tabular}{|c|c|c|}
\hline
$J$ & $M=1,\Lambda=1$ & $M=2,\Lambda=3$    \\
\hline
-0.04 & 1.84$\cdot10^{-5}$ & 1.71$\cdot10^{-5}$     \\
-0.06 & 1.68$\cdot10^{-4}$ & 1.56$\cdot10^{-4}$     \\
-0.1  & 1.39$\cdot10^{-3}$ & 1.43$\cdot10^{-3}$       \\
-0.15 & 5.32$\cdot10^{-3}$ & 5.81$\cdot10^{-3}$        \\
-0.2  & 1.19$\cdot10^{-2}$ & 1.33$\cdot10^{-2}$         \\
 \hline
 \end{tabular}
\end{center}
 \label{Tabl:TK}
\end{table}

\begin{table}[htb]
\caption{
$\chi_{\mathrm{loc}}(0)$  with varying $J$; $M=1$,  extrapolation to $\Lambda=1$}
\begin{center}
\begin{tabular}{|c|c|c|c|}
\hline
$J$ & -0.06 & -0.1 &  -0.2    \\
\hline $\chi_{\mathrm{loc}}(0)$ & 454.488 & 52.9607  &  5.93006         \\
 \hline
 \end{tabular}
\end{center}
 \label{Tabl:chi_loc_0}
\end{table}

The behavior of ${\chi _{\mathrm{loc}}(T)}$ at low temperatures is shown in Fig.~\ref{Fig_chi_loc_TL}.
The empirical linear dependence
\begin{equation}
\frac{\chi _{\mathrm{loc}}(T)}{\chi _{\mathrm{loc}}(0)} = 1 + \frac{2}{3} {\chi _{\mathrm{loc}}(0)} T
\label{Eq:chi_loc_1ch}
\end{equation}
describes the low-temperature behavior rather well, with $\chi_{\mathrm{loc}}(0)$ presented in Table~\ref{Tabl:chi_loc_0}.

Generally, the low temperature dependence of $\chi_{\mathrm{loc}}$ is  rather unusual: instead of nearly constant value
below $T_K$ one observes a linear behavior. To confirm that this  behavior is not an artifact, NRG calculations were
performed for a series of values of $\Lambda$ (see Fig.~\ref{Fig_chi_loc_1chan}).

Thus the situation in the singular DOS case  differs from that in the flat band case where typical parabolic Fermi liquid
dependence $T_{K}^{\mathrm{loc}}\chi_{\mathrm{loc}} (T)=0.097 - 0.07 (T/T_{K}^\mathrm{loc})^{2}$ takes place (see
Fig.~\ref{Fig_chi_loc_1chan_parabola}).

If the singularity is slightly shifted  from the Fermi level, with $\mu$ being the value of the shift, we should have the
crossover to the typical for the flat-band temperature behavior. However at small $\mu$ the dependence of $\chi
_{\mathrm{loc}}(T)$ acquires at lowering temperature  a linear increase instead of a maximum (see inset in Fig.~\ref{Fig_chi_loc_1chan}).

For completeness we introduce  an alternatively defined susceptibility $\chi_{\mathrm{imp}}$, which can be expressed as a
difference of magnetic susceptibilities of the whole system and the system without impurity:
\begin{equation}
\chi _{\mathrm{imp}}(T)=\chi _{\mathrm{tot}}(T)-\chi _{\mathrm{band}}(T)\ ,
\label{chi_imp}
\end{equation}%
where $\chi _{\mathrm{tot}}$ is the total magnetic susceptibility, and $\chi _{\mathrm{band}}$ is the susceptibility of
non-interacting band electrons. Since in this definition magnetic field acts on the whole system, this quantity can be
experimentally determined from magnetic measurements; it is also usually treated in Bethe ansatz solutions. In the
flat-band case we have at low temperatures the standard Fermi-liquid behavior $T_K^{\mathrm{imp}}\chi _{\mathrm{imp}}(T)=
0.103 - 0.11(T/T_K^{\mathrm{imp}})^{2}$, where according to Wilson~\cite{Wilson} we use the definition
$T_{K}^{\mathrm{imp}}\chi_{\mathrm{imp}}(T_{K}^{\mathrm{imp}})=0.0701$.

As discussed, e.g., in Refs.~\cite{Santoro} and~\cite{Cox} (Sect.6.2.2), $\chi_{\mathrm{imp}}$
and $\chi_{\mathrm{loc}}$
can be different. In our case $\chi_{\mathrm{imp}}(T)$ at not too low $T$ demonstrates the usual universal Kondo behavior
with $T_{K}^{\mathrm{imp}}$ being defined according to Wilson.

\begin{figure}[htbp]
\includegraphics[width=0.8\columnwidth, angle=0]{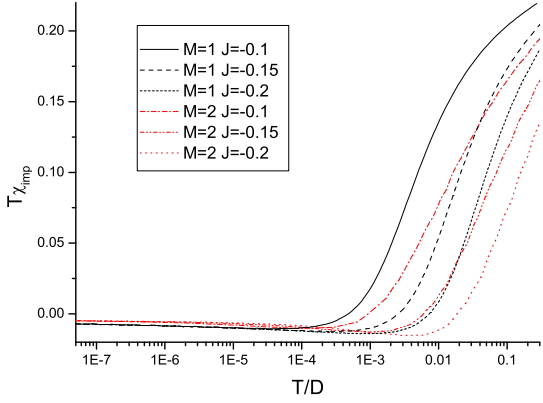}
\caption{NRG results for $T\chi_{\mathrm{imp}}$. One-channel ($M=1$) and two-channel ($M=2$) cases with singular DOS. } \label{Fig:chi_imp}
\end{figure}

However, with decreasing temperature $\chi_{\mathrm{imp}}(T)$ deviates
from the universal behavior and changes its sign.  At very low $T$ we have irrespective of $J$
(Fig.~\ref{Fig:chi_imp})
\begin{equation}
T\chi _{\mathrm{imp}}(T)\approx\frac{-0.115}{\ln(D/T)} \ .
\label{eq:chi_nrg}
\end{equation}
The temperature of loss of universality for $\chi _{\mathrm{imp}}$ corresponds to that for $\chi _{\mathrm{loc}}(T)$: both
the sign change in the former and maximum in the latter are connected with overcompensation of impurity spin by conduction
electrons. For the case $M=1$, $\chi _{\mathrm{imp}} $ was discussed in details in Ref.~\cite{Zhur2009}.

To compare these results with above perturbative consideration, we remember that in our case,
as follows from the structure of perturbation theory,
$ g_{ef}(T)\rightarrow G_{ef}(T)= g_{ef}(T)\ln(D/T)$, and
\begin{equation}
G^*- G_{ef} \sim 1/\ln(D/T)
\label{sc4}
\end{equation}
according to~(\ref{sc3}). Thus the terms, proportional to $G^*- G_{ef}$, should give $1/\ln(D/T)$-contribution to
thermodynamic characteristics. In particular, the $1/\ln(D/T)$ terms in $\chi_{\mathrm{imp}}(T)$ are expected. However,
the calculation of $\chi_{\mathrm{imp}}(T)$ requires a careful collection of all the contributions to susceptibility.

A similar behavior is obtained for the impurity entropy (Fig.~\ref{Fig:entropy_combine})
and for specific heat,
\begin{equation}
S_{\mathrm{imp}}(T)\approx \frac{-1.3}{\ln(D/T)}, \
C_{\mathrm{imp}}(T)\approx\frac{-1.3}{\ln^2 (D/T)} \ .
\label{Eq:SpecHeat}
\end{equation}
According to Ref.~\cite{Gan} (Eq.44), the entropy in the flat band multichannel model contains the contribution $(\pi^2/4)
(M g_{ef}^3(T )- (3/8) M^2 g^4_{ef}(T))$ with $g^*- g_{ef}(T) \sim T^\Delta$. To leading order in $1/M$, this yields the
contribution, proportional to $ [g^*- g_{ef}(T)]^2 \sim T^{2\Delta} $.
One can expect that in our case terms linear in $g_{ef}(T)\rightarrow G_{ef}(T)= g_{ef}(T)\ln(D/T)$
will occur in the entropy. However, to obtain a correct description of strong coupling regime,
a more accurate analysis is required.

%
These results correct somewhat our previous calculations~\cite{Zhur2009}. The occurrence of the inverse-logarithm
contributions is in a qualitative agreement with the above scaling consideration.

The physical picture can be explained as follows. At low temperatures the impurity spin is completely screened and we come
to the situation where $\chi_{\mathrm{imp}}$ is determined by the contribution of conduction electrons which is
independent of $J$. The effective bandwidth in the singular case decreases, and the situation is close to
that in the model with a hole at the magnetic impurity site ($J\rightarrow -\infty $): here the
total magnetic susceptibility of the system is smaller than that of bare electrons,
so that the contribution $\chi_{\mathrm{imp}}$ is evidently negative.

\begin{figure}[htbp]
\includegraphics[width=0.8\columnwidth, angle=0]{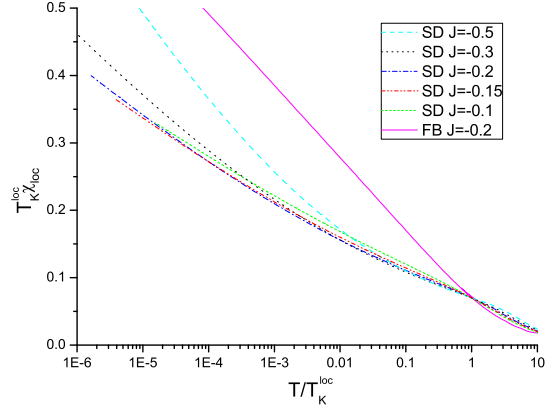}
\caption{NRG results for $\chi _{\mathrm{loc}}$. The case of the two-channel Kondo
model with (SD) and without (FB) logarithmic
singularity in one-electron DOS.} \label{Fig:chi_loc_2chan}
\end{figure}

Thus in the strong coupling regime (at low temperatures)  the problem is reduced to the resonant level model.
A quantitative analytical solution in this model can be obtained similar to \cite{Mitchell}.
In terms of the impurity free energy one obtains

\begin{equation}
S_{\mathrm{imp}} = -\frac{\partial F_{imp}}{\partial T}
             = -\sum_{\sigma} \int_{-\infty}^{\infty} \frac{\mathrm{d}z}{\pi}
                        \frac{z}{\cosh^{2}z}  \Im \ln (-G_{r,\sigma}^{-1}(2Tz))
\label{eq:S_imp}
\end{equation}

\begin{equation}
T\chi_{\mathrm{imp}}= -\frac{1}{4}\sum_{\sigma} \int_{-\infty}^{\infty} \frac{\mathrm{d}z}{\pi}
                          \frac{\sinh z}{\cosh^{3} z} \Im \ln (-G_{r,\sigma}^{-1}(2Tz))
\label{eq:Chi_imp}
\end{equation}
where $G_{r,\sigma}(z)$  is the retarded Green's function of the resonant level hybridized with the band.

Calculating the
 Green's function of the level hybridized with the singular DOS we have
\begin{equation}
G_{r,\sigma}^{-1}(z)= z+i0^{+}-v^{2}G_{r,\sigma}^{0}(z)
\end{equation}
where $v$ is an effective hybridization matrix element and
\begin{equation}
G_{r,\sigma}^{0}(z)=
                \rho\left[P\int_{-D}^{D}\mathrm{d}\epsilon
                           \frac{\ln|4D/\epsilon|}{z-\epsilon} -
                           i\pi\ln|4D/z|
                           \theta (D^2-z^2) \right]
\end{equation}
is the band Green's function for the singular DOS at the resonant level site, $\theta(x)$ is the Heaviside step function. Since the low $T$
behavior is dominated by small $|z|\ll D$  one has
\begin{equation}
 \Im \ln (-G_{r,\sigma}^{-1}(z))
                           \approx -\frac{\pi}{2} +
                                    \mathrm{sign}(z)\arctan\frac{\pi}{2\ln|4D/z|}
\label{eq:argln}
\end{equation}
and we derive for leading corrections irrespective of $v$
\begin{equation}
 S_{\mathrm{imp}} \approx -\frac{\ln 4}{\ln |D/T|}, \quad
 T\chi_{\mathrm{imp}} \approx -\frac{1/8}{\ln|D/T|}
\label{RLR}
\end{equation}%
in a fair agreement with the NRG results (\ref{eq:chi_nrg}),(\ref{Eq:SpecHeat}).
The above consideration shows that the unusual low temperature behavior is solely determined by the
logarithmic van Hove singularity of the band spectrum (cf.~\cite{Mitchell}),
the singular contribution having essentially one-electron nature.

Now we pass to the two-channel situation. In the flat-band case  $\chi_{\mathrm{loc}} (T)$ is known to behave as $\ln(T_{K}/T)$~\cite{Cox}
(such a behavior was also reproduced by our test calculation, Fig.~\ref{Fig:chi_loc_2chan}). However,
for the logarithmic DOS  the local susceptibility $\chi _{\mathrm{loc}}(T)$ demonstrates a power-law  non-Fermi-liquid
behavior (Fig.~\ref{Fig:chi_loc_2chan}), which can be fitted at low $T$ as
\begin{equation}
T_{K}^{\mathrm{loc}}\chi _{\mathrm{loc}}(T) \sim (T_{K}^{\mathrm{loc}}/T)^{\alpha}
\label{2ch}
\end{equation}%
where  $\alpha$  slightly decreases with decreasing $|J|$ (see Table~\ref{Tabl:alpha}).

\begin{table}[htb]
\caption{Estimated values for the exponent $\alpha$  in~(\ref{2ch}). Note that the accuracy at small $|J|$ is low}
\begin{center}
\begin{tabular}{|c|c|c|c|c|c|}
\hline
$J$      & -0.5  &  -0.3 & -0.2 & -0.15 & -0.1         \\
\hline
$\alpha$ & 0.19  &  0.15 & 0.14 &  0.12 &  0.1         \\
 \hline
 \end{tabular}
\end{center}
 \label{Tabl:alpha}
\end{table}

Thus the power-law behavior of physical quantities occurs in the presence of the DOS singularity already for $M=2$ (in the flat-band case this takes place starting from $M=3$).
One can suppose that the singularity leads to an effective increase of the number of scattering channels.
Although it is often difficult to distinguish between the
experimental logarithmic and power-law dependences, especially in the case of small exponents, this conclusion is important
from the theoretical point of view.

As demonstrated in Ref.~\cite{Gan} from the $1/M$-expansion, in the flat-band case the change in the gyromagnetic ratio
(which enters total magnetic susceptibility) leads to a change in numerical factors only, so that  both
$\chi_{\mathrm{loc}}(T)$ and  $\chi_{\mathrm{imp}}(T)$ are positive and behave in  similar way. However, in our case the
behavior of $\chi_{\mathrm{imp}}(T)$ is again qualitatively different: we have the dependence
(see Fig.~\ref{Fig:chi_imp})
\begin{equation}
T\chi _{\mathrm{imp}}(T)\approx\frac{-0.075}{\ln(D/T)} \ .
\end{equation}

For the $M=2$ flat-band Kondo model the ground state impurity entropy is known to be equal to
$\ln2/2$~\cite{Cox}. Our NRG
calculations confirm this value (see Fig.~\ref{Fig:entropy_combine}). For the singular DOS the temperature behavior is
again unusual: at low $T$ the entropy approaches this value from below according to the law
\begin{equation}
S_{\mathrm{imp}}(T)\approx \frac{\ln2}{2}-\frac{1.3}{\ln(D/T)}
\approx S_{\mathrm{imp}}^{1channel} + \frac{\ln2}{2} \ .
\label{eq:2chentropy}
\end{equation}
Thus the temperature dependence in (\ref{eq:2chentropy}) is the same as in the resonant level model (\ref{RLR}).

We demonstrated that the behavior in the Kondo model with singular logarithmic DOS differs radically from that in the
smooth DOS model. This has unusual behavior with negative impurity magnetic susceptibility and entropy at low $T$. On the
other hand, the local magnetic susceptibility which is determined by the linear response remains positive. This has a
shallow maximum in the one-channel case and demonstrates a power-law NFL behavior for two-channel case. As for the
$1/M$-expansion, this yields the results which differ from those for smooth DOS case by the replacement of the effective
coupling constant by $G_{\mathrm ef}$. However, the NRG method enables one to obtain a more detailed information (although
the calculations are rather cumbersome because of slow convergence in the singular case). In particular, NRG calculations
reproduce the Bethe ansatz results for the  flat band $M=2$ case (logarithmic factors in local magnetic susceptibility). In
the case of singular DOS, the  $M=2$ results turn out to be different (weak power-law divergence). We note also absence of
Wilson's self-similarity in terms of $T/T_K$. When shifting the singularity from the Fermi level, the system can demonstrate
non-trivial crossovers at lowering $T$, which can be observed experimentally.

The research was carried out within the state assignment of FASO of Russia (theme ``Quantum'' No. 01201463332) and supported in part by Ural Branch of Russian Academy of Science (project no. 15-8-2-10).


\end{document}